\documentclass[a4paper]{article}
\usepackage{listings}
\usepackage{xcolor}
\usepackage{multirow}
\usepackage{jheppub}  
\usepackage{dcolumn}
\usepackage[english]{babel}
\usepackage[utf8x]{inputenc}
\usepackage[T1]{fontenc}
\usepackage{palatino}
\usepackage{lineno}
\usepackage{amsmath}
\usepackage{afterpage}
\usepackage{graphicx, stackengine, subcaption}
\usepackage[colorinlistoftodos]{todonotes}
\usepackage[colorlinks=true]{hyperref}
\usepackage{makeidx}
\newcommand{\be}{\begin{equation}}  
\newcommand{\ee}{\end{equation}}  
\newcommand{\bea}{\begin{eqnarray}}  
\newcommand{\eea}{\end{eqnarray}}  

\usepackage{ptdr-definitions}
\usepackage{hepunits}
\usepackage{heppennames2}

\newcommand{\rootsNN} {\ensuremath{\sqrt{\smash[b]{s_{_{\mathrm{NN}}}}}}\xspace}
\newcommand{\roots} {\ensuremath{\sqrt{\smash[b]{s}}}\xspace}
\newcommand{\PbPb}{\ensuremath{\mathrm{Pb}\mathrm{Pb}}\xspace}
\newcommand{\pPb}{\ensuremath{\Pp\mathrm{Pb}}\xspace}
\newcommand{\pp}{\ensuremath{\Pp\Pp}\xspace}

\newcommand{\mub}{\ensuremath{\,\mu\text{b}}\xspace}
\newcommand{\xthree}{\PGccDoP{3872}}
\newcommand{\stt}{\ensuremath{\sigma_{\ttbar}}\xspace}
\newcommand{\stttwol}{\ensuremath{2.56\pm 0.82}}
\newcommand{\stttwoljets}{\ensuremath{2.02\pm 0.69}}

\makeindex
\begin{document}

\vspace*{1.2cm}

\thispagestyle{empty}
\begin{center}
{\LARGE \bf Review of results using heavy ion collisions at CMS}

\par\vspace*{7mm}\par

{

\bigskip

\large \bf \href{https://gkrintir.web.cern.ch}{Georgios Konstantinos Krintiras} \\ (on behalf of the CMS Collaboration)}

\bigskip

{\large \bf  E-Mail: gkrintir@cern.ch}

\bigskip

{The University of Kansas}

\bigskip

{\it Presented at the Workshop of QCD and Forward Physics at the EIC, the LHC, and Cosmic Ray Physics in Guanajuato, Mexico, November 18--21 2019}

\bigskip

{\it Published in \href{https://doi.org/10.17161/1808.30727}{doi.org/10.17161/1808.30727}}

\vspace*{15mm}

{  \bf  Abstract }

\end{center}
\vspace*{1mm}

\begin{abstract} 
{U}Ultrarelativistic heavy ion collisions at the laboratory provide a unique chance to study quantum chromodynamics (QCD) under extreme temperature (${\approx}150\,\MeV$) and density (${\approx}1\,\GeV/\fm^3$) conditions. Over  the  past  decade,  experimental results  from  LHC have shown further evidence for the formation of the quark-gluon plasma (QGP),  a phase that is thought to permeate the early Universe and is formed in the high-density neutron-star cores.  Various QCD predictions that model the behavior of the low-$x$ gluon nuclear density, a poorly explored region, are also tested. Since the photon flux per ion scales as the square of the emitting electric charge $Z^2$, cross sections of so far elusive photon-induced processes are extremely enhanced as compared to nucleon-nucleon collisions. Here, we review recent progress on CMS measurements of particle production with large transverse momentum or mass, photon-initiated processes, jet-induced  medium  response, and heavy quark production. These high-precision data, along with novel approaches, offer stringent constraints on initial state, QGP formation and transport parameters, and even parametrizations beyond the standard model. 

\end{abstract}
  
\section{Introduction}
\label{Intro}

 The Compact Muon Solenoid (CMS) is a "general purpose" detector, however, equally well suited for the study of heavy ion collisions at LHC~\cite{Chatrchyan:2008zzk}. Since the first lead-lead (PbPb) collisions recorded at CMS in 2010, and after almost ten years of operation, a wealth of measurements are available for understanding hadron and nuclear "static" (\eg, mass generation and spectra) and "dynamic" (\eg, cross sections) properties.  Initially, heavy ion collisions were proposed to study basic features of quantum chromodynamics (QCD) matter via its excitation to phases where quarks and gluons are no more confined into hadrons. Although high-density regimes of QCD are routinely formed in the laboratory using nucleus-nucleus collisions~\cite{Bazavov:2018mes}, and might exist in the present Universe~\cite{Weih:2019xvw}, we witness signatures for their existence in "small-systems", \eg, proton-proton  (\pp)~\cite{Khachatryan:2010gv} as well as proton-nucleus~\cite{CMS:2012qk} collisions. Although in the former case the physical origin of "long-range correlations", \ie, two-particle angular correlations with large pseudorapidity gap, is interpreted as a consequence of hydrodynamic expansion of the produced medium with initial-state fluctuations, the underlying  mechanisms are not yet understood in \pp and proton-nucleus collisions~\cite{Shen:2020gef}.  Collisions of point-like objects, \eg, electron-positron~\cite{Badea:2019vey}, can serve as reference to the observed long-range correlations in the small systems.
 
 The extracted properties of the quark-gluon plasma (QGP), created in the extreme environment of high temperature (${\approx}150\,\MeV$) and energy density (${\approx}1\,\GeV/\fm^3$), signify an almost ideal liquid with short lifetime (${\approx}10\,\fm$) and large opacity against the partons traversing it~\cite{Sirunyan:2018ktu}. The QGP response strongly depends on the geometrical overlap ("centrality") of heavy ion collisions.
 "Central" collisions, at small impact parameter $b$, yield large and round interaction regions, whereas peripheral  collisions,  characterized  by  larger  values  of  $b$, result in smaller interaction regions with more pronounced spatial anisotropy.  The centrality dependence of  various observables  provides,  then,  insight  into  their dependence on global geometry. Instances with no overlap, \ie, $b$ being larger than twice the radius of the nuclei, are well suited to study photon (\PGg)-mediated interactions: "ultraperipheral" collisions (UPC) are the energy frontier for the photoproduction of  heavy  quark  and  antiquark bound  states ("quarkonia"), and "dijets", \ie, pairs consisting of the most ("leading") and the second most ("subleading") energetic jets, reaching at LHC $\PGg\Pp$ ($\PGg\PGg$) center of mass energies \roots significantly higher than at HERA (LEP)~\cite{Klein:2020fmr}. Here, we briefly present the latest results from the CMS Collaboration on "quark matter and beyond", \ie, including novel measurements and beyond the standard model (BSM) searches that are competitive with, or at least complementary to, \pp\ studies.

\section{Heavy ion collisions at LHC}
\label{sec:performance}

\begin{figure}[ht!]
\begin{center}
\includegraphics[width=.5\textwidth]{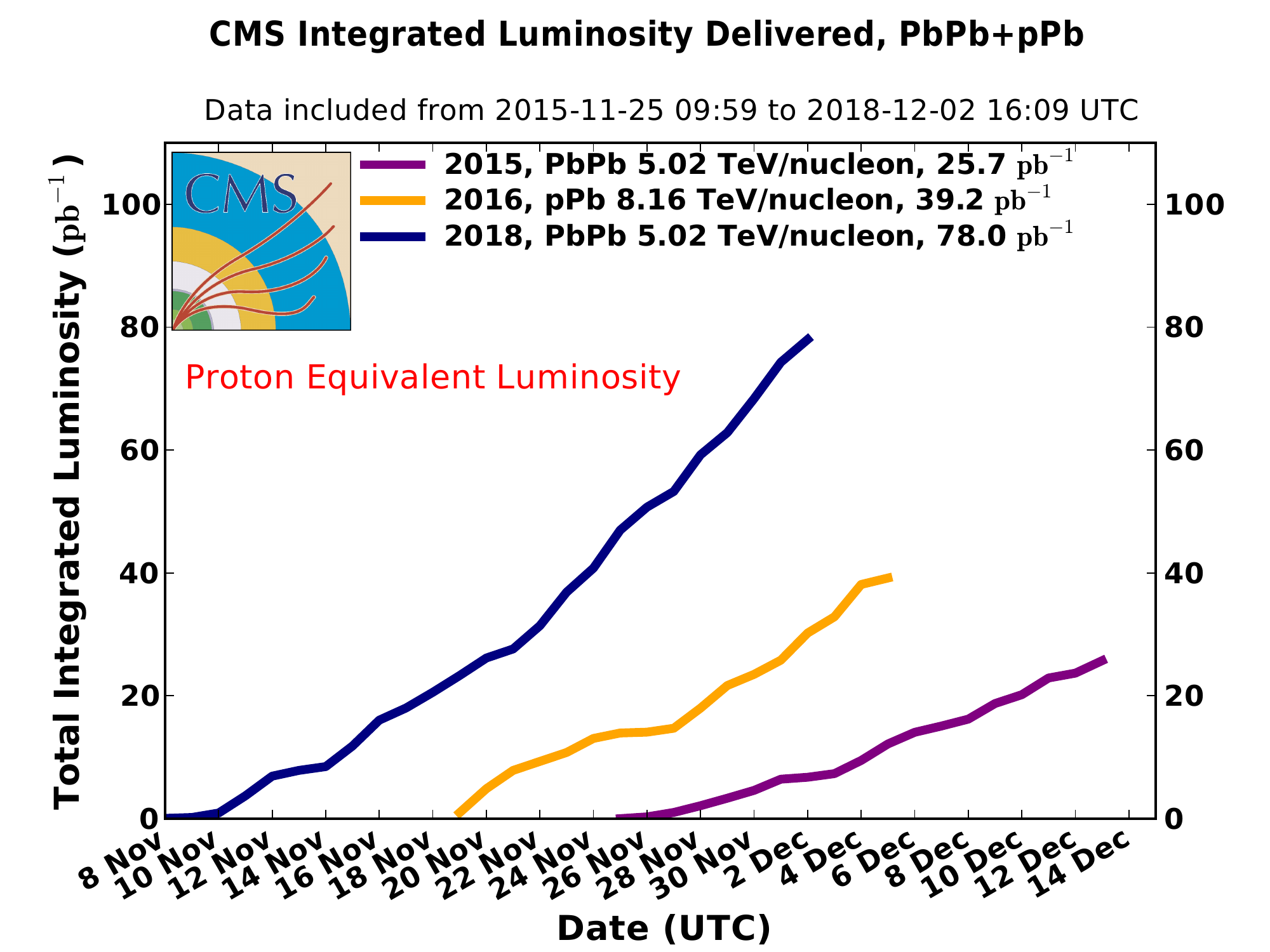}
\end{center}
\caption{\label{fig:Lumi} The $\Lumi_\mathrm{NN}$-integrated luminosity delivered to CMS for \PbPb\ and \pPb\ collisions in Run 2~\cite{CMS:2018fkg}.}
\end{figure}

Four LHC experiments recorded data with heavy ion collisions during "Run 2" (2015--2018), rendering the so-called "fair luminosity sharing" among them challenging. Physics runs have also been carried out with proton-lead (\pPb) collisions, a mode of operation~\cite{Salgado:2011wc} that was not initially foreseen~\cite{Jowett:2006au}. The standard LHC operation includes heavy ion or "reference" \pp\ collisions at the same nucleon-nucleon \roots energy (\rootsNN) during roughly one month per year. The performance until the end of Run 2 has been excellent, reaching instantaneous luminosities of about six (eight) times higher than the design (physics-case) value of $10^{27} (10^{29})\,\percms$ in PbPb (\Pp{}Pb), equivalent to a nucleon-nucleon luminosity of $\Lumi_\mathrm{NN}\sim 10^{32}\percms$. The $\Lumi_\mathrm{NN}$-integrated luminosity delivered to CMS~\cite{CMS:2018fkg} is shown in Fig.~\ref{fig:Lumi} for \PbPb\ and \pPb\ collisions.

The excellent performance was made possible through a series of improvements in the LHC and the injector chain. For the next \PbPb\ run in 2021, it is planned to further increase the total LHC beam intensity through a decrease of bunch spacing to 50\,\unit{ns}{}, resulting to a total of 1\,232 bunches in the ring. 
However, any increase of ion luminosity is intrinsically limited by the risk of superconductivity loss in the magnets. As mitigation, it is therefore planned to install additional collimators during the second long shutdown (2018--2021) to allow for higher luminosity and intensity~\cite{Bruce:2014sya, Lechner:2014cya, ApollinariG.:2017ojx}.
Using the predicted beam and machine configuration~\cite{Citron:2018lsq}, the future luminosity performance during one-month runs is estimated about 3 and 700\nbinv for \PbPb\ and \pPb, respectively. 

\section{Hard probes and photon-induced processes}
\label{sec:hardprobes}

The proton structure at high momentum-transfer $Q^2$, as encoded in  the  collinearly  factorized  parton  distribution  functions (PDFs), enters the weighted product with high-energy ("hard") parton-parton scattering cross sections. While the latter are process specific and are computed in perturbative QCD (pQCD) at different levels of accuracy, PDFs are deemed universal functions of $Q^2$ and Bjorken-$x$ obtained by the well-established means of global analyses ("fits")  using hard-process  data. Likewise, their counterparts for nucleons bound in nuclei, \ie, the nuclear PDFs (nPDFs), are essential in studying the production of hard probes in QGP.  As such, the uncertainty in nPDFs mainly stems from the available data, and hence the lack of constraints in certain phase space regions. 

\begin{figure}[ht!]
\begin{center}
\includegraphics[width=1.\textwidth]{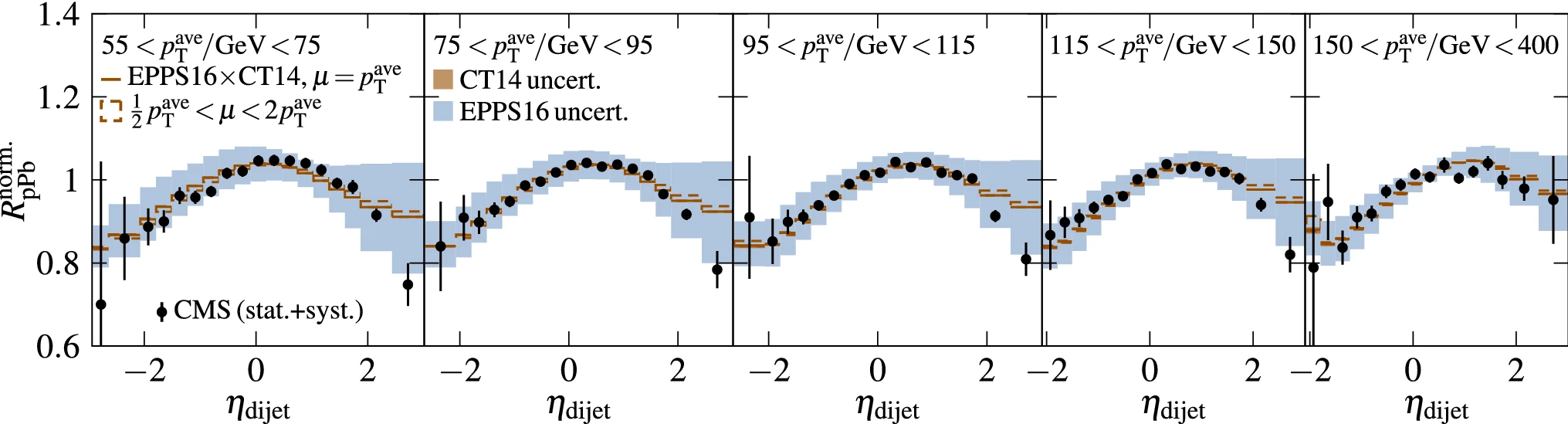}
\end{center}
\caption{\label{fig:dijet_RpPb} The nuclear modification factor between \pPb and \pp dijet pseudorapidity differential cross sections. Black markers show the data from CMS~\cite{Sirunyan:2018qel} with vertical bars showing the statistical and systematical uncertainties added in quadrature. Solid lines represent the pQCD calculation with average \pt scale choice using the central nPDF set of CT14~\cite{Dulat:2015mca} and EPPS16~\cite{Eskola:2016oht}.
Figure extracted from Ref.~\cite{Eskola:2018sxu}.
}
\end{figure}

A recent example is the measurement of dijet pseudorapidity ($\eta_{\text{dijet}}$) spectra in \pp and \pPb collisions at 5.02\,\TeV~\cite{Sirunyan:2018qel}, where the uncertainty in the nuclear modification factor $R_{\pPb}$ between the \pPb and \pp spectra appears significantly smaller than predictions with various nPDFs (Fig.~\ref{fig:dijet_RpPb} from Ref.~\cite{Eskola:2018sxu}). Processes involving electroweak gauge (\eg, \PW or \PZ) bosons and top quarks are also powerful probes of the light quark and gluon nPDFs, respectively. The production of W bosons~\cite{Sirunyan:2019dox} and top quarks~\cite{Sirunyan:2017xku} is studied in \pPb\ collisions at $\rootsNN = 8.16$\,\TeV. In the former case, the results already favor PDF calculations that include nuclear modifications, and provide constraints on the nPDF global fits. In the latter case, although the top quark is a novel and theoretically precise probe of nPDFs due to its high mass~\cite{Sirunyan:2019jyn}, the measured cross section of its pair (\ttbar) production, \stt, is still consistent with the expectations from scaled pp data. The exploration of parton densities in nuclei in a broad ($x$, $Q^2$) kinematic range~\cite{CMS:2018qpj} is a priority for the high-luminosity \PbPb\ and \pPb\ physics programs (see Section~\ref{sec:performance}).

\begin{figure}[ht!]
\begin{center}
\includegraphics[width=.42\textwidth]{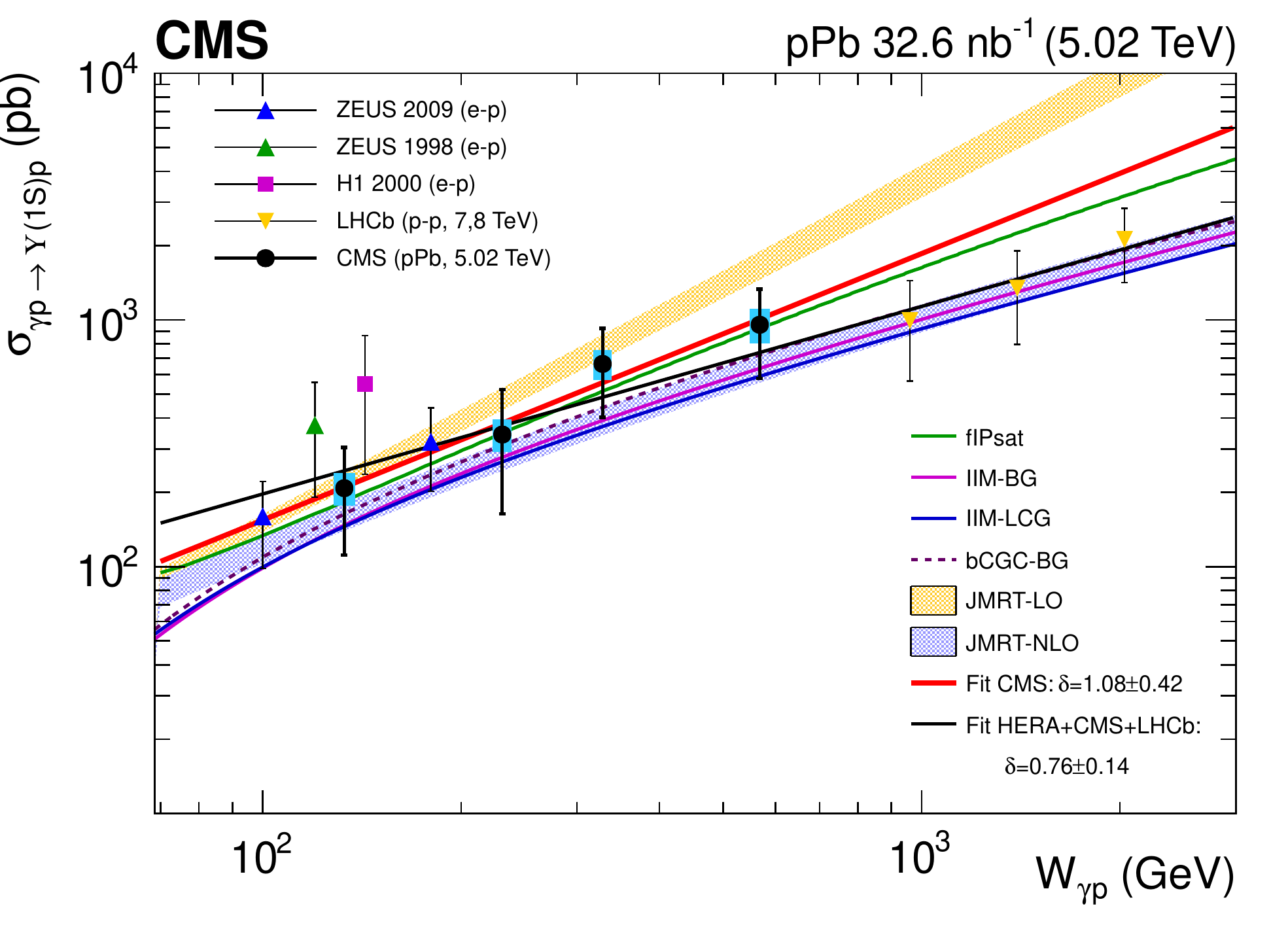}
\includegraphics[width=.48\textwidth]{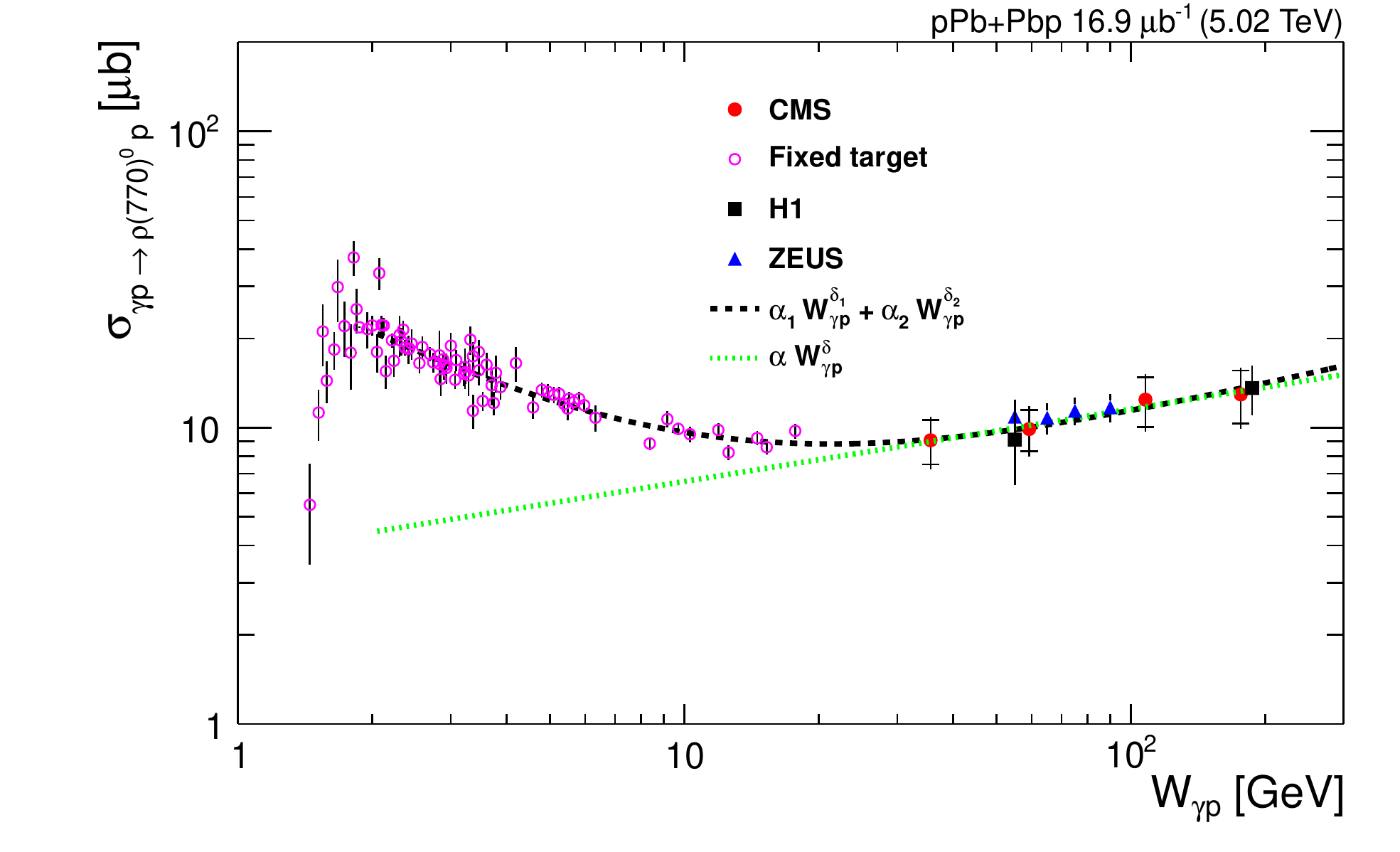}
\end{center}
\caption{\label{fig:VM} Cross section for the exclusive $\PgU\mathrm{(1S)}$~\cite{Sirunyan:2018sav}  (left) and $\Pgr^0$~\cite{Sirunyan:2019nog} (right)  photoproduction as a function of $W_{\PGg\Pp}$ compared to previous data as well as to various theoretical predictions or theory-inspired fits. The bars show the statistical uncertainty, while the boxes or outer bars represent the systematic uncertainty or the statistical and systematic uncertainties added in quadrature. }
\end{figure}

Ultrarelativistic heavy ions are accompanied by strong electromagnetic fields; the latter can be treated as a flux of quasi-real photons that scale with the square of the emitting electric charge $Z^2$ thus their radiation from $\mathrm{Pb}$ ions is strongly enhanced compared to the \Pp case. Quasi-real photons can fluctuate into a quark-antiquark pair,  which can then turn into a vector meson ($\mathrm{VM}$) upon interacting with the other nucleon in UPC. In particular, "exclusive" $\mathrm{VM}$ photoproduction, $\PGg\Pp\to{\mathrm{VM}}\Pp$, bridge previously unexplored regions of parton fractional momenta from the HERA measurements, \eg, $x\approx10^{\text{-4}}$--$10^{\text{-2}}$ in the case of Ref.~\cite{Sirunyan:2018sav} where $\mathrm{VM}\equiv \PgU\mathrm{(nS)}$ (with $n=1,2,3$), meaning UPC can be used in the same way as electron-proton collisions. The incoming nucleons remain intact after the interaction and only the VM is produced, with the process referred to as "exclusive". Also, in these events, contrary to symmetric colliding systems, one can determine the $\PGg$ direction and hence the $\PGg\Pp$ centre-of-mass energy, $W_{\PGg\Pp}$, unambiguously. The data, within their currently large statistical uncertainty, are consistent with various pQCD approaches that model the behavior of the low-$x$ gluon density (Fig.~\ref{fig:VM}, left) and provide new insights on the gluon distribution in the proton in this poorly explored region. Some models suggest that the energy dependence of the integrated cross section may provide evidence of gluon saturation, as investigated in Ref.~\cite{Sirunyan:2019nog} where the exclusive $\Pgr^0$ photoproduction is  measured,  for  the  first  time,  in UPC \pPb\ collisions at $\rootsNN=5.02$\,\TeV\ (Fig.~\ref{fig:VM}, right).

\begin{figure}[ht!]
\begin{center}
\includegraphics[width=.48\textwidth]{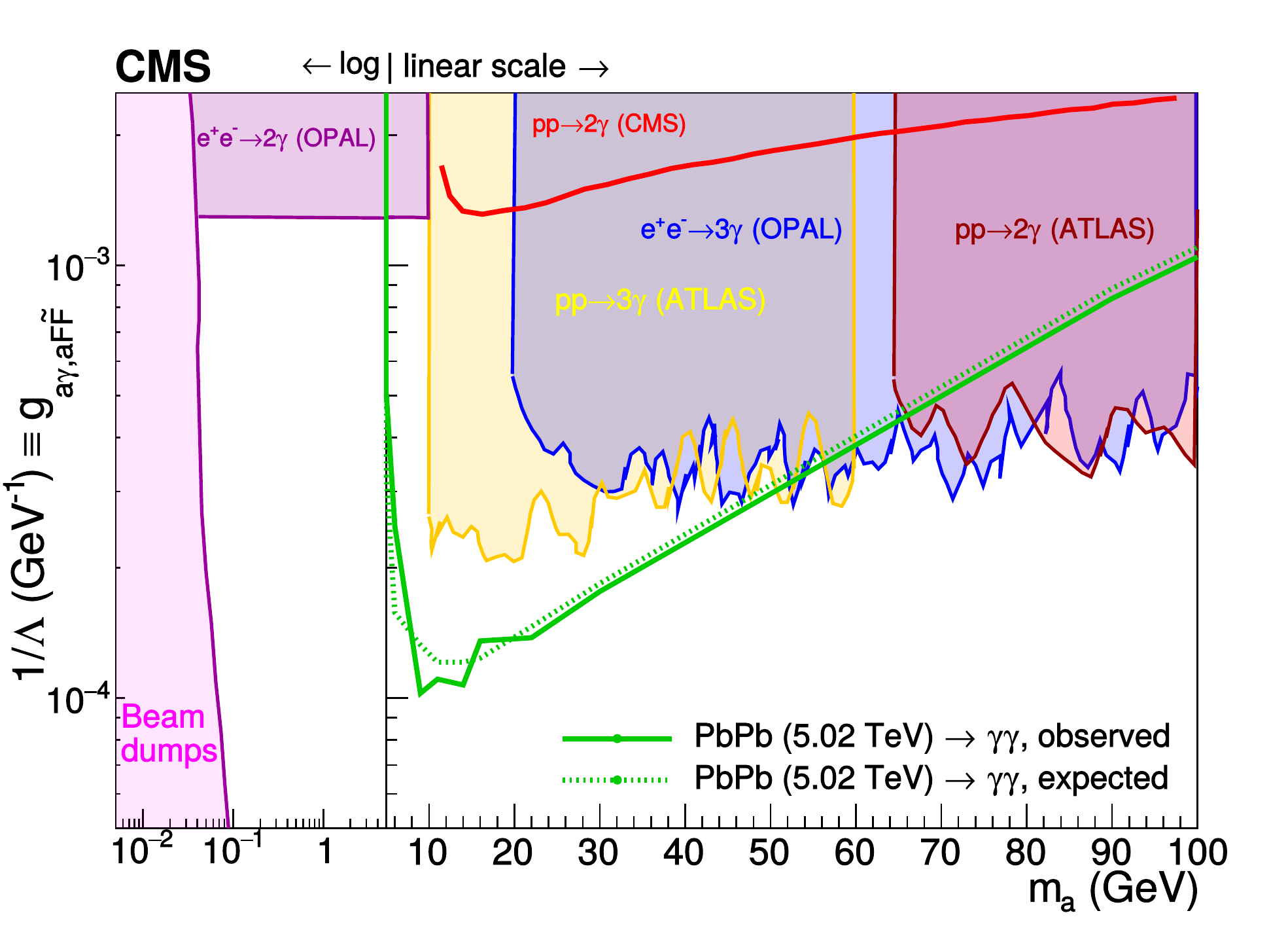}
\includegraphics[width=.48\textwidth]{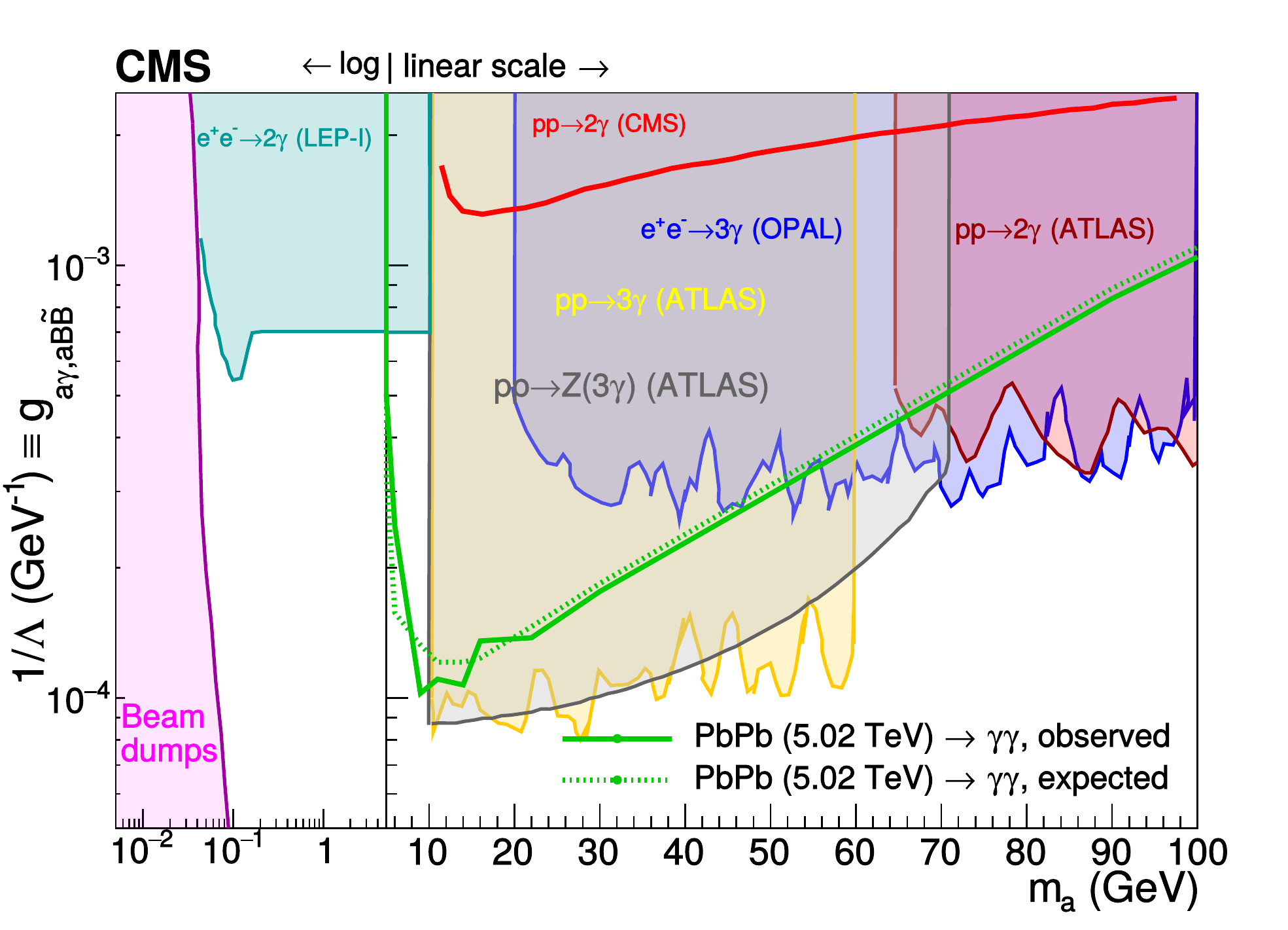}
\end{center}
\caption{\label{fig:ALPs} Exclusion limits at 95\% CL in the $g_{\text{a}\PGg}$--$m_{\text{a}}$ plane, assuming ALP coupling to \PGg only (left) and including the hypercharge coupling, hence involving the \PZ boson (right), from previous measurements and compared to the present limits using \PbPb\ collisions~\cite{Sirunyan:2018fhl}.}
\end{figure}

Even accounting for roughly 100 times lower instantaneous luminosity than \pp collisions, several BSM searches appear more competitive in nuclear than \pp collisions~\cite{Bruce:2018yzs}. For instance, in case that BSM is manifested with low couplings to the SM and at relatively low masses, experimental conditions related to heavy ion collisions---with almost vanishing pileup, optimal primary vertex reconstruction, low-\pt thresholds applied to online filters, and ``clean'' exclusive final states in UPC---present relative merits compared to \pp\ studies. A characteristic example is Ref.~\cite{Sirunyan:2018fhl} where the measured exclusive \PGg{}\PGg invariant mass distribution is used to search for narrow resonances such as pseudoscalar axion-like (a) particles (ALPs) produced in the process $\PGg\PGg \to \text{a} \to \PGg\PGg$. Exclusion limits at 95\% confidence level (CL) are set on the $\PGg\PGg \to \text{a} \to \PGg\PGg$ cross section for ALPs with masses $m_{\text{a}}=5$--90 \GeV. The cross section limits are then used to set exclusion limits in the two-dimensional plane of the ALP coupling to photons $g_{\text{a}PGg}\equiv1/\Lambda$ (with $\Lambda$ being the BSM energy scale) and $m_{\text{a}}$. For ALPs coupling to the electromagnetic (and electroweak) current, the derived exclusion limits are currently the best over the $m_{\text{a}}=5$--50  (5--10)\,\GeV\ mass range, as shown in Fig.~\ref{fig:ALPs}.

\section{Jet modifications}
\label{sec:Jet_mods}

High-momentum partons are produced by hard scatterings that occur over a timescale $\tau\sim\pt^{\mathrm{-1}}<10\,\fm$ thus are expected to undergo "energy loss" as they traverse the QGP. The mechanisms by which the partons distribute their energy to the medium (radiative or collisional energy loss) as well as the color dependence (\eg, energy loss due to the different color charges, coherent or incoherent gluon  radiation, etc), are still not fully understood. The particles resulting from the parton fragmentation and hadronization are clustered into jets (of cone size $R=\sqrt{\left(\Delta y\right)^2+\left(\Delta \phi\right)^2}$) that are used as parton proxies to examine the QGP properties. Parton energy loss manifests itself in various experimental observables, including the suppression $R_\mathrm{AA}$ of high-\pt hadrons~\cite{Sirunyan:2018ktu} and jets, including its system size dependence~\cite{Sirunyan:2018eqi}, as well as modifications of the jet properties (\eg,  dijet momentum balance~\cite{Chatrchyan:2012nia}, charged particle number densities, jet fragmentation functions, and jet shape~\cite{Sirunyan:2018qec,Sirunyan:2018jqr,Sirunyan:2018ncy,Sirunyan:2019dow}) and parton shower~\cite{CMS-PAS-HIN-18-020}. These phenomena are collectively referred to as "jet quenching" that can be related  to  the  transport  and  thermodynamic  QGP properties.

\begin{figure}[ht!]
\begin{center}
\includegraphics[width=.68\textwidth]{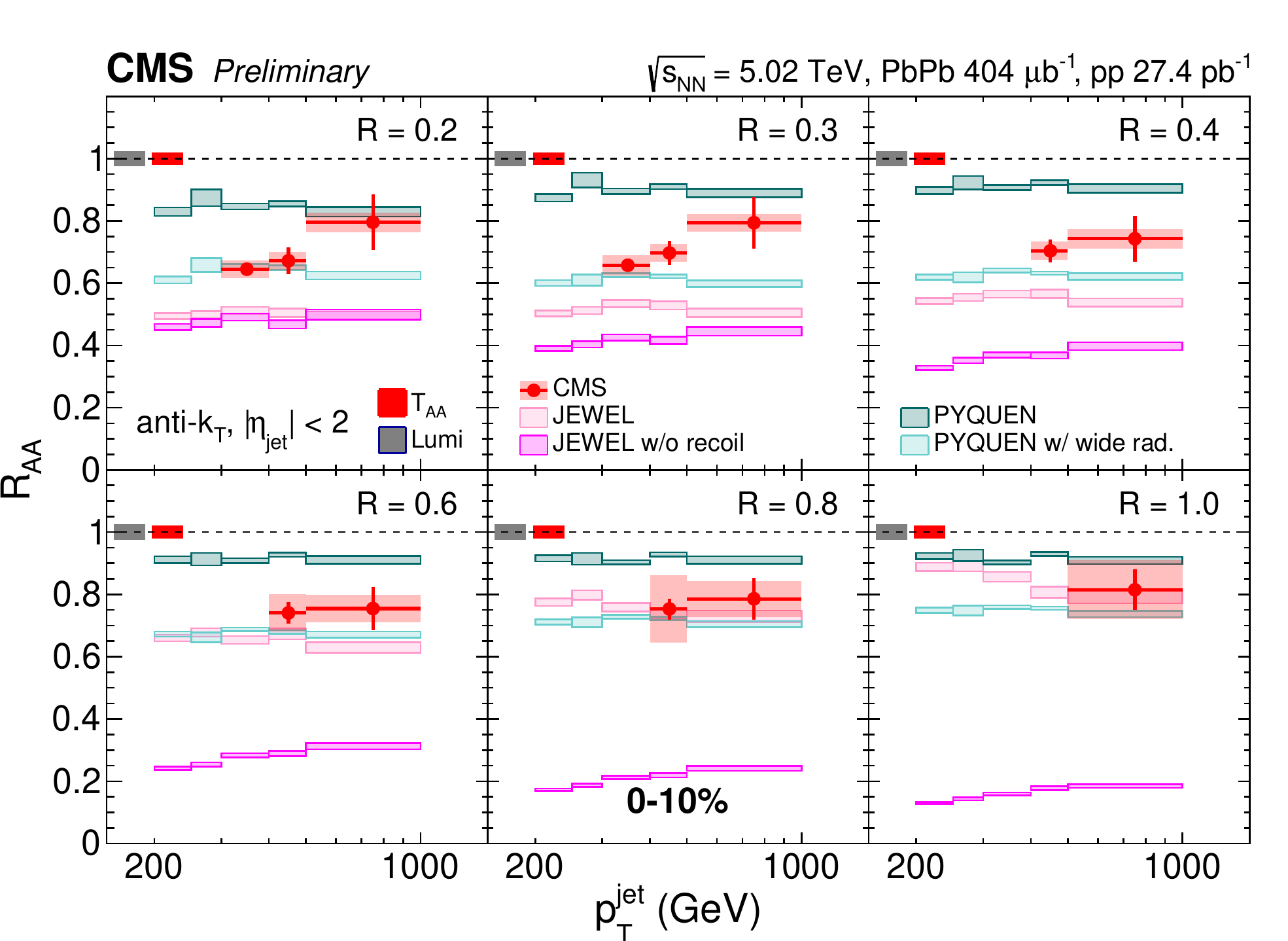}
\end{center}
\caption{\label{fig:RAA} The QGP-induced modification on jets as a function of jet \pt for various $R$ and 0--10\% centrality class~\cite{CMS-PAS-HIN-18-014}. The statistical uncertainty is represented as a vertical line, while the systematic uncertainty is shown as a shaded box. The uncertainty due to luminosity and  nuclear overlap function for \pp\ and \PbPb\ collisions, respectively, are shown as colored boxes on the dashed line at 1. Data are compared to predictions from Jewel~\cite{Zapp:2013vla} (orange and purple) and \textsc{pyquen}~\cite{Lokhtin:2005px} (teal and green) generators.}
\end{figure}

Recently, and for the first time, strong suppression of high-\pt large-$R$ jets is observed in the 0--10\% most central collisions~\cite{CMS-PAS-HIN-18-014}. Whereas the various predictions from quenched jet event generators, theoretical models, and analytical calculations grasp reasonably the $R_\mathrm{AA}$ \pt evolution for jets reconstructed with $R \lesssim 0.4$, they produce a less uniform description of the QGP-induced behavior of jets at larger $R$ (Fig.~\ref{fig:RAA}). Another observation, which does not support previous interpretations based on color-charge-dependent jet quenching, is achieved with a template-fitting method using the "jet charge" observable~\cite{Sirunyan:2020qvi}. Jet charge, defined as the momentum-weighted sum of the electric charges of particles inside a jet, is sensitive to the electric charge of the particle initiating a parton shower and can be used to discriminate between gluon- and quark-initiated jets. No evidence is seen for a significant decrease (increase) in gluon- (quark-) like prevalence in a sample of high-\pt jets in \PbPb\ collisions. In contrast to the \PbPb\ and xenon-xenon systems, in \pPb\ collisions no suppression is observed in the low-\pt region, whereas a weak momentum dependence is seen for $\pt>10$\,\GeV, leading to a moderate excess above unity~\cite{Khachatryan:2016odn}.

\section{Heavy quark dynamics}
\label{sec:Heavy_dynamics}

The majority of secondary particle production in heavy ion collisions ends up as collective, thermalized bulk QCD matter which is well described using almost ideal relativistic fluid dynamics. The QGP behavior is largely governed by an equation of state that exhibits a phase transition from confined matter with hadronic degrees of freedom below the transition and deconfined partonic  matter above~\cite{Andronic:2017pug}. The long-range two- or higher-order particle (with large $\eta$ gap) and near-side azimuthal correlations constitute an effective tool to probe these QGP properties. These "radial" or "anisotropic flow" correlations~\cite{Ollitrault:1993ba,Voloshin:1994mz,Poskanzer:1998yz} are typically parametrized by coefficients in a Fourier expansion, $v_n$ ($n\geq 1$), and can provide information about the initial collision geometry (\eg, the "elliptic" flow harmonic, $v_2$) and its fluctuations (\eg, the "triangular" flow harmonic, $v_3$).

Recent measurements~\cite{CMS-PAS-HIN-19-008} of prompt \PDz\ (\PaDz) meson $v_2$ and $v_3$ in \PbPb\ collisions at \rootsNN\ = 5.02\,\TeV, as a function of \pt, rapidity and centrality, extend the \pt coverage up to ${\sim}60$\,\GeV\ and provide more differential information. Motivated by the search for a strong electric field possibly created in \PbPb\ collisions, the
first measurement of the $v_2$ difference ($\Delta v_2$) between \PDz\ and \PaDz as a function
of rapidity is studied. The rapidity-averaged $v_2$ difference is measured
$\left\langle\Delta v_2 \right\rangle = 0.001 \pm 0.001\stat \pm 0.003\syst$. No effect of electric field on charm hadron collective flow is thus observed, within the experimental uncertainty, and future model comparisons can provide constraints on the QGP electric conductivity~\cite{Gursoy:2018yai}.

\begin{figure}[h]
 \begin{center}
\includegraphics[width=0.37\textwidth]{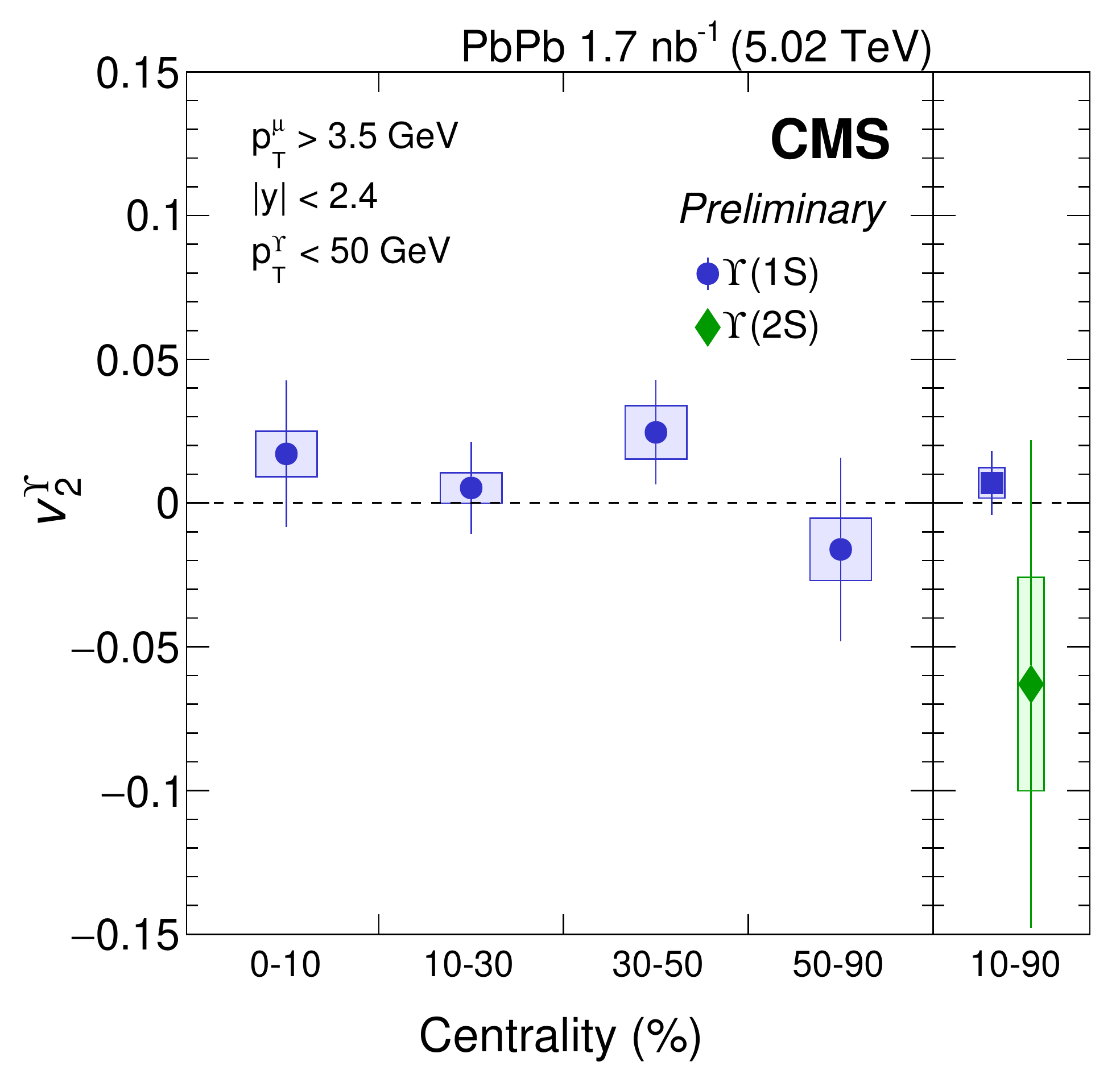}
\includegraphics[width=0.45\textwidth]{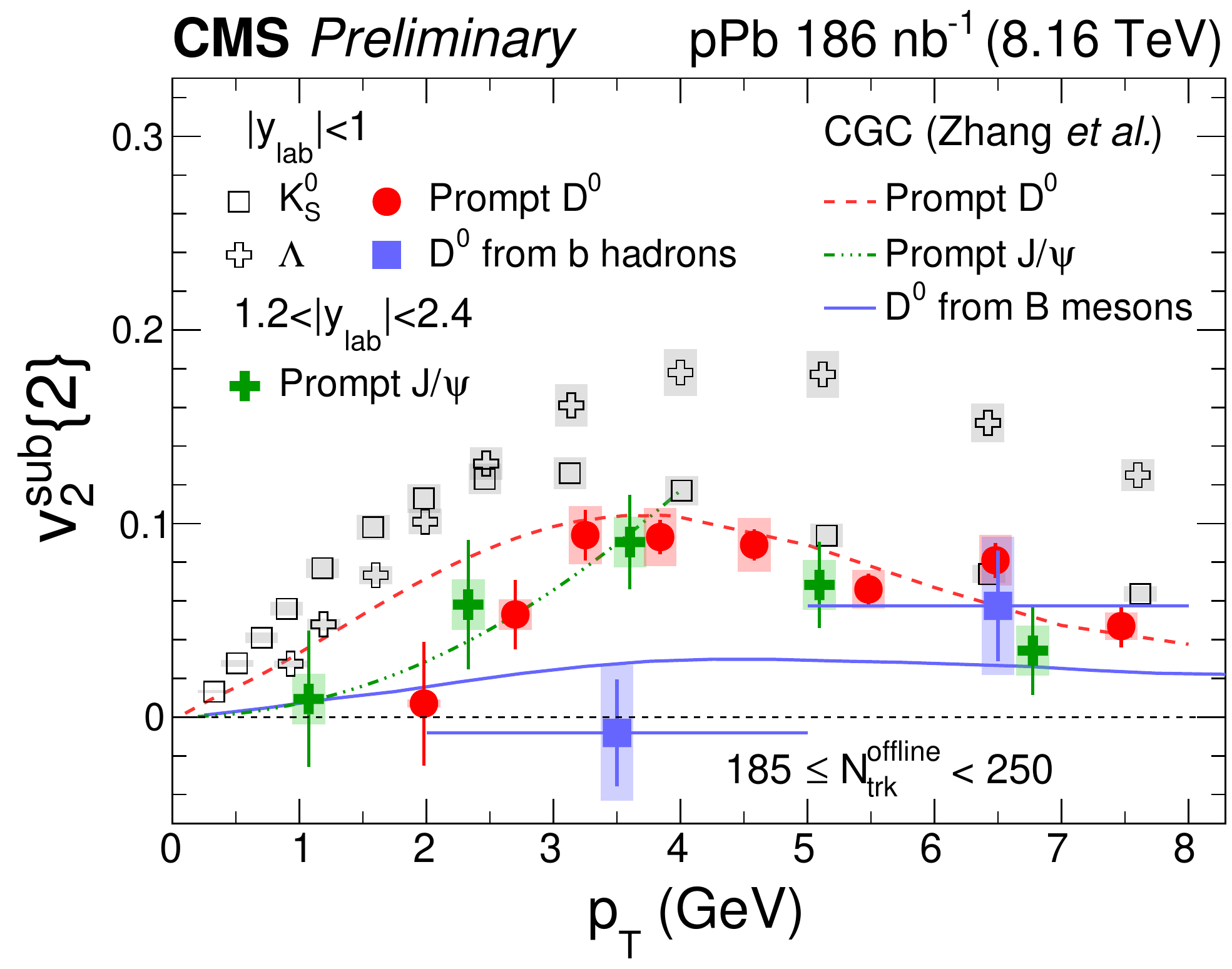}
\caption{
(Left) The \pt-integrated $v_2$ values for $\PgU\mathrm{(1S)}$ and $\PgU\mathrm{(2S)}$ mesons measured in \PbPb\ collisions $\rootsNN  = 5.02$\,\TeV\ in the 0--90\% centrality range~\cite{CMS-PAS-HIN-19-002}. The error bars (bands) denote the statistical (systematic) uncertainty. (Right) Results of $v_2$ for prompt and nonprompt \PDz\ mesons~\cite{CMS-PAS-HIN-19-009}, as well as \PKzS, \PgL, and \JPsi, as a function of \pt in \pPb\ collisions at $\rootsNN  = 8.16$\,\TeV~\cite{Sirunyan:2018toe,Sirunyan:2018kiz}.
The error bars (shaded areas)  correspond to the statistical (systematic) uncertainty. Lines show the theoretical calculations under the color glass condensate framework of Ref.~\cite{Zhang:2019dth}. }
   \label{fig:heavy_quarks}
 \end{center}
\end{figure}

In the case of quarkonia states with different binding energies, their azimuthal dependence, which is largely independent of nPDFs, can reflect the extent of the "screening", \ie, at what level their binding energy is weakened by the surrounding partons, hence revealing the QGP thermal environment. 
The $v_2$ values for $\PgU\mathrm{(1S)}$ and, for the first time, $\PgU\mathrm{(2S)}$ mesons are measured (Fig.~\ref{fig:heavy_quarks}, left) and found to be consistent with zero over the kinematic range studied~\cite{CMS-PAS-HIN-19-002}, contrasting with the measured \JPsi results in \PbPb\ collisions~\cite{Khachatryan:2016ypw}, and suggesting different QGP effects for charmonia and bottomonia. Because of different contribution of regeneration between $\PgU\mathrm{(1S)}$ and $\PgU\mathrm{(2S)}$ meson production, this measurement additionally provides new inputs to the production mechanisms of bottomonia, complementing the sequential suppression pattern already seen for $\PgU\mathrm{(nS)}$ (with $n=1,2,3$) mesons~\cite{Sirunyan:2018nsz}.

Heavy flavor quark collectivity is also seen in small-system collisions~\cite{CMS-PAS-HIN-19-009}, measuring $v_2>0$ for prompt \PDz\ in \pp\ collisions---comparable to light-flavor hadron species---and extracting a mass dependence of heavy flavor hadron $v_2$ in \pPb\ collisions, including, for the first time, open beauty hadrons via nonprompt \PDz\ mesons (Fig.~\ref{fig:heavy_quarks}, right) . 

\section{New probes}
\label{sec:New_probes}

\begin{figure}[h]
 \begin{center}
\includegraphics[width=0.43\textwidth]{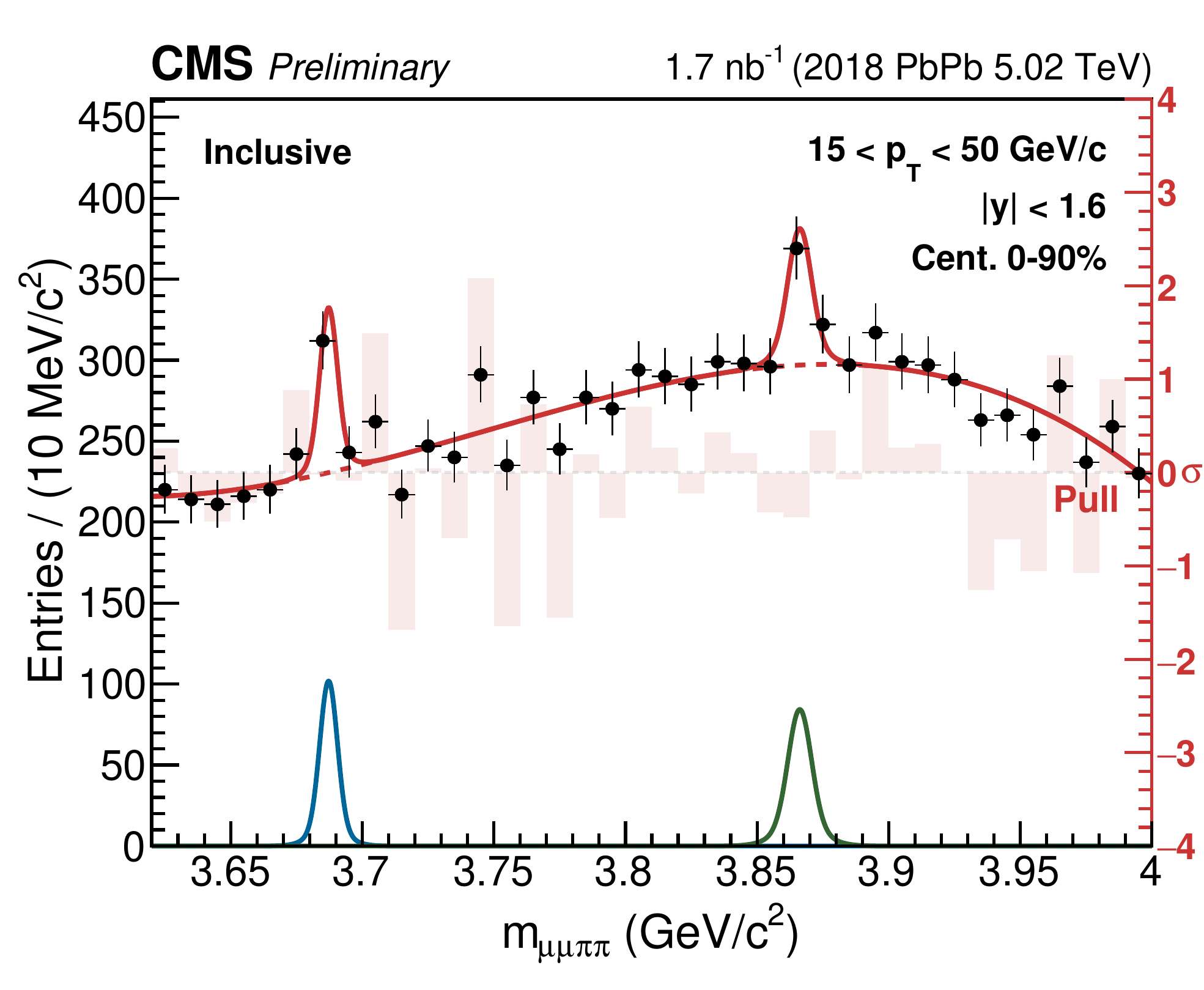}
\includegraphics[width=0.45\textwidth]{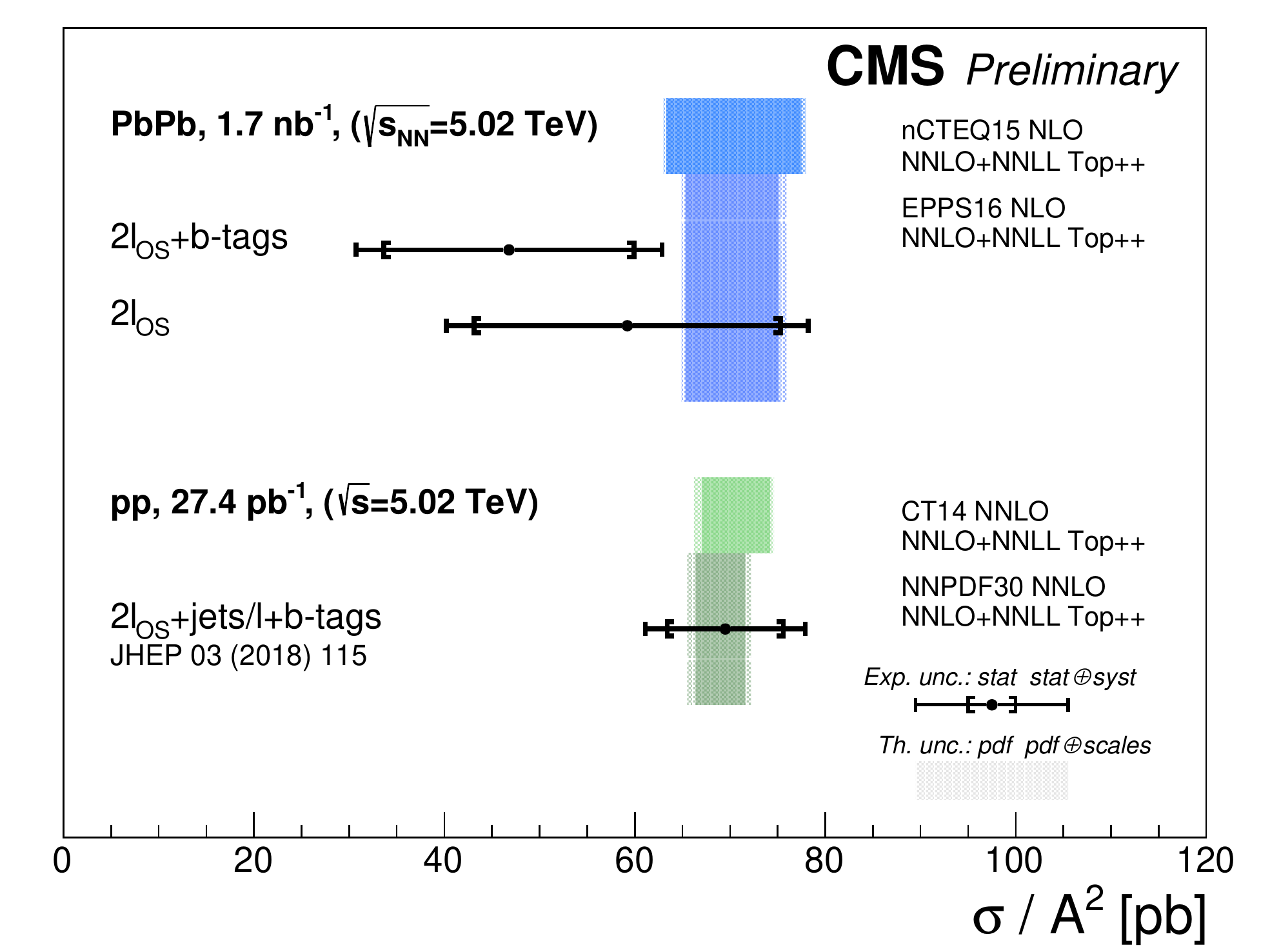}
\caption{
(Left) Invariant mass distribution $m_{\mu^+\mu^- \Pgpp\Pgpm}$ in \PbPb\ collisions at $\rootsNN  = 5.02$\,\TeV~\cite{CMS-PAS-HIN-19-005}, denoting the production of $\psi(\text{2S})$ (blue) and \xthree (green) particles. The red line presents the unbinned maximum-likelihood fit result, with the "pull" distribution represented by the red boxes. (Right) Inclusive \ttbar\ cross sections measured with two methods in dilepton final states in \PbPb\ collisions at $\rootsNN=5.02$\,\TeV\ (scaled by $A^{\text{-2}}$)~\cite{CMS-PAS-HIN-19-001}, and pp results at $\roots=5.02$\,\TeV~\cite{Sirunyan:2017ule}. The measurements are compared with theory predictions at higher-order accuracy in QCD~\cite{Czakon:2013goa}. The inner (outer) experimental uncertainty bars include statistical (statistical and systematic, added in quadrature) uncertainties. The inner (outer) theory uncertainty bands correspond to nuclear~\cite{Eskola:2016oht,Kusina:2016fxy} or free-nucleon~\cite{Dulat:2015mca,Ball:2014} PDF (PDF and scale, added in quadrature) uncertainties. }
   \label{fig:new_probes}
 \end{center}
\end{figure}

The multi-\TeV\ energies available at the LHC heavy ion collisions have opened up the possibility
to measure, for the first time, various exotic mesons and high-mass elementary particles. The $X(\text{3872})$, also known as \xthree, is such an exotic meson, first observed by the BELLE Collaboration~\cite{Choi:2003ue} and subsequently studied at electron-position and hadron colliders (most recently in Ref.~\cite{Chatrchyan:2013cld}), whose nature is still not fully understood. In the QGP, the production of the \xthree state can be enhanced or depleted depending on the spatial configuration of the exotic state. The recent measurement of the inclusive prompt \xthree production~\cite{CMS-PAS-HIN-19-005}, here reconstructed via the decay chain $\xthree \to \JPsi\, \Pgpp\Pgpm\to \mu^+\mu^- \Pgpp\Pgpm$ (Fig.~\ref{fig:new_probes}, left), could also provide a new test on the statistical hadronization mechanism~\cite{Andronic:2017pug}, a remarkably successful phenomenological description of the yields of 
"stable" (with respect to strong interactions) hadrons in central relativistic heavy ion collisions. 

The characteristic feature of experimental signatures is their sensitivity to initial- or final-state effects
integrated over the QGP lifetime, the latter increasing as a function of \rootsNN and the atomic mass $A$ of the ions being collided. At variance with measurements considered so far in the literature, top quark, a colored particle that decays mostly within the QGP, provides a novel way to study differentially the space-time evolution of the QGP, hence offering the opportunity to resolve the QGP and "unveiling its yoctosecond structure"~\cite{Apolinario:2019vqt}. We demonstrate~\cite{CMS-PAS-HIN-19-001}, for the first time, that top quark decay products are identified, irrespective of whether interacting with the medium (bottom quarks) or not (leptonically decaying \PW bosons). Dilepton final states are selected, and the \ttbar\ cross section is measured from a likelihood fit to a multivariate discriminator using lepton kinematic variables. The measurement is additionally performed considering the jets originating from the hadronization of bottom quarks, which improve the sensitivity to the signal process. The measurements, $\stt= \stttwoljets$ and $\stttwol$\mub, consistent with each other and the expectations from scaled proton-proton data as well as perturbative QCD (Fig.~\ref{fig:new_probes}, right), constitute the first step towards using the top quark as a novel tool to probe the QGP.

\section{Summary}

The quark-gluon plasma  (QGP)  is  a state  of transient nuclear matter in which composite quantum chromodynamics (QCD) states loose their identity and dissolve into a nearly ideal, strongly interacting fluid of quarks and gluons. The existence of QGP was proposed already in the mid-seventies after it was realized that asymptotic freedom in QCD predicts force weakening at short distances. Therefore, one of the most challenging questions in nuclear physics  is  to  identify the QGP structure  and  its phases. 

Despite the successful description provided by the QCD Lagrangian, our knowledge away from the perturbative limit is still limited. The aim of ultrarelativistic heavy ion collisions is to bridge this gap and to contribute to the understanding of thermodynamics and collective QCD phenomena. Hard probes and photon-initiated interactions serve in this context to provide information and constraints for cold and hot nuclear matter effects which cannot be obtained by studying bulk QCD matter alone. Imaging the QGP formation and evolution via initial- and final-state interactions of produced and outgoing partons lies at the heart of nuclear PDF and tomography studies, respectively. Altogether these measurements contribute to comprehensive modeling of all aspects of the dynamics of heavy ion collisions.

Some of the features of the QGP are the strong collective anisotropic flow and high opacity to jets. Collective flow is observed, among others, as the mass-dependent transverse momentum (\pt) spectra of light or heavy particles, while parton energy loss comes out as the suppression in the production of high-\pt particles. Recent measurements in high-multiplicity proton-proton and proton-nucleus collisions revealed flow-like patterns, and along with the nuclear modification factors, already covering \pt ranges up to the \TeV\ scale, can provide stringent constraints on cold and hot nuclear matter effects.

With the advent of the LHC, the energy reach for ultraperipheral collisions extended significantly, including studies on nuclear structure and modifications, and even searches for beyond the standard model signatures. Additional measurements like the evidence of exotic meson and top quark production demonstrate the versatility of the CMS experiment, and provide novel probes of the locally deconfined state with a lifetime of a few fm. The future opportunities for high-density QCD studies with ion and proton beams at the LHC are unprecedented given the enlarged per month integrated luminosity. 

\section*{Acknowledgements}

The work is supported in whole by the Nuclear Physics (NP) program of the U.S. Department of Energy (DOE) with number \href{https://pamspublic.science.energy.gov/WebPAMSExternal/Interface/Common/ViewPublicAbstract.aspx?rv=d1ddcae6-235b-4163-ae34-01fce58e5f90&rtc=24&PRoleId=10}{DOE DE-SC0019389}.

\bibliographystyle{auto_generated.bst} 
\bibliography{references.bib}

\printindex

\end{document}